\documentclass[10pt,conference]{IEEEtran}
\IEEEoverridecommandlockouts
\usepackage[numbers,sort&compress]{natbib}
\usepackage{rotating}
\usepackage{array}
\newcolumntype{L}[1]{>{\raggedright\let\newline\\\arraybackslash\hspace{0pt}}m{#1}}
\newcolumntype{C}[1]{>{\centering\let\newline\\\arraybackslash\hspace{0pt}}m{#1}}
\newcolumntype{R}[1]{>{\raggedleft\let\newline\\\arraybackslash\hspace{0pt}}m{#1}}
\usepackage{amsmath,amssymb,amsthm,amsfonts,bm}
\usepackage{algorithmicx,algorithm}
\usepackage{tablefootnote}
\usepackage{algpseudocode}
\usepackage{threeparttable}
\algnewcommand{\algorithmicand}{\textbf{ and }}
\algnewcommand{\algorithmicor}{\textbf{ or }}
\algnewcommand{\OR}{\algorithmicor}
\usepackage{graphicx}
\usepackage{multirow}
\usepackage{url}
\usepackage{soul}
\usepackage{textcomp}
\usepackage{xcolor}
\def\BibTeX{{\rm B\kern-.05em{\sc i\kern-.025em b}\kern-.08em
    T\kern-.1667em\lower.7ex\hbox{E}\kern-.125emX}}
\newcommand{\paolo}[1]{\textcolor{purple}{ #1}}
\newcommand{\ck}[1]{\textcolor{magenta}{ #1}}
\newcommand{\ff}[1]{\textcolor{blue}{ #1}}
\newcommand{\newff}[1]{\textcolor{blue}{ #1}}

\algdef{SE}[SUBALG]{Indent}{EndIndent}{}{\algorithmicend\ }%
\algtext*{Indent}
\algtext*{EndIndent}

\usepackage{subfigure}
\usepackage[utf8]{inputenc}
\usepackage{mathtools}
\begin{document}
\renewcommand\paolo[1]{#1}
\renewcommand\ck[1]{#1}
\renewcommand\ff[1]{#1}
\renewcommand\newff[1]{#1}

\title{
Measuring Data Similarity for 
Efficient \\ Federated Learning: A Feasibility Study
\thanks{The present work has been funded by the European Union’s Horizon
2020 Marie Skłodowska Curie Innovative Training Network Greenedge (GA.
No. 953775).}}

\author{\IEEEauthorblockN{Fernanda Famá\IEEEauthorrefmark{1}\IEEEauthorrefmark{2}, Charalampos Kalalas\IEEEauthorrefmark{1}, Sandra Lagen\IEEEauthorrefmark{1}, Paolo Dini\IEEEauthorrefmark{1}}
\IEEEauthorblockA{{\IEEEauthorrefmark{1}Centre Tecnològic de Telecomunicacions de Catalunya (CTTC/CERCA)}\\
\IEEEauthorblockA{\IEEEauthorrefmark{2}\ff{Dept. of Signal Theory and Communications, Universitat Politècnica de Catalunya, Spain}}
Email: \{fernanda.fama, ckalalas, slagen, pdini\}@cttc.es}}

\maketitle

\begin{abstract}
\ck{In} multiple federated learning 
\ck{schemes}, a random subset of clients sends \ck{in each round} their model updates to the server for aggregation.
Although this client selection \ck{strategy} \ck{aims to reduce} 
communication overhead, it \ck{remains} 
energy and computationally inefficient, especially \ck{when} considering resource-constrained devices \ck{as clients. This is because conventional random client selection overlooks the content of exchanged information and falls short of providing} 
a mechanism to reduce the transmission of \ck{semantically} redundant \ck{data.} 
To overcome this challenge, we propose clustering the clients 
\ck{with the aid of} similarity metrics, where 
\ck{a single} client from each of \ck{the formed} 
clusters is selected in each round to participate in the federated training. To 
evaluate our approach, \ck{we perform an extensive feasibility study} considering \ck{the use of} nine statistical metrics in the clustering \ck{process}. \ck{Simulation} results \ck{reveal} that, when considering a scenario \ck{with high data heterogeneity of clients}, 
\ck{similarity-based clustering can reduce the number of required rounds compared to the baseline random client selection. In addition, energy consumption can be notably reduced from
23.93\% to 41.61\%, 
for those similarity metrics with an equivalent number of
clients per round as the baseline random scheme.}
\end{abstract}

\begin{IEEEkeywords}
federated learning, similarity metrics, clustering, client selection
\end{IEEEkeywords}

\vspace{-0.1cm}

\section{Introduction}

The edge computing paradigm has become a breakthrough solution that allows certain processing resources to be allocated close to the data sources rather than relying only on cloud computing centers. 
However, 
\ck{edge-computing} devices have limited \ck{computational and energy} resources, which makes training complex machine learning (ML) models challenging. Recently, federated learning (FL) has emerged as an alternative for performing ML tasks in a distributed and privacy-preserving manner by having multiple clients collaborate without sharing their data\newff{~\cite{mcmahan2017communication}}. This is because local learning clients share their parameters with a central server, which aggregates them to create a proper global model after a series of rounds.



Despite the privacy and computational sharing benefits of FL across devices, the use of multiple clients can create a communication bottleneck~\cite{shahid2021communication}.
\ck{This may be further exacerbated by
the energy consumption spent training FL models on such resource-constrained devices}. 
Therefore, several studies have \ck{proposed} 
techniques to select, \paolo{at every round}, a subset of clients to participate in training the global model. 
However, deciding how many and which clients to \ck{choose} is not straightforward, \ck{therefore} a random \ck{selection method} 
is usually adopted. 
\ck{In turn, this may inadvertently}
lead to the selection of similar and redundant information for aggregation, which requires a higher commitment of resources to obtain a global model suitable for all clients\newff{~\cite{balakrishnan2022diverse}}. \ck{Random selection may thus have}
a negative impact on learning efficiency, fairness, convergence \paolo{and, eventually, energy consumption}. 
Apart from this,  data on client devices may not be evenly distributed in 
\ck{real-world FL tasks},
and non-iid solutions need to be considered. \ck{The distribution divergence of non-iid data introduces significant challenges in FL~\cite{ma2022state}}  
such as lower accuracy, delays in model communication, and \ck{slower} 
model convergence.

Motivated
by the 
challenges \ck{mentioned above}
, we \ck{hereby advocate for} 
using similarity metrics \newff{\cite{visus2021taxonomy}}
\ck{as a means of}
grouping 
FL clients with similar information. To achieve this, we consider scenarios with a skewed label distribution, 
\ck{and we perform cluster formulation with the aid of the following similarity metrics:}
cosine function; mean squared error; Euclidean, Manhattan, and Chebyshev distances; maximum mean discrepancy; Kullback-Leibler and Jensen-Shannon divergences; and Wasserstein distance. 
\ck{The incorporation of such metrics} 
allows us to 
\ck{leverage} correlations in the \ck{local} information to 
form \textit{semantically informative} clusters.
Selecting clients from these clusters minimizes the transmission of redundant information \ck{
by harnessing the underlying heterogeneity of
local training datasets.}
Consequently, it accelerates FL training efficiently. 
\ck{Our twofold} contribution 
\ck{can be summarized}
as follows:
\begin{itemize}
    \item We explore the \ck{feasibility of applying} 
    different \ck{statistical} similarity metrics \ck{on local data to perform informative client clustering. The incorporation of similarity metrics in client selection allows FL training to be performed in a way that promotes dissimilarity of selected clients.
    Our goal resides in not only reducing}
    redundancy in the FL training phase, \ck{but also in quantifying the potential energy-efficiency gains.}
    \item We 
    \ck{perform an elaborate comparison between similarity-based clustering and random client selection with the aid of multiple FL performance indicators,} 
    such as accuracy, number of rounds, and energy consumption, \ck{for different degrees of skewness in label distributions.} \ck{Key
observations stemming from the trade-off between the number of rounds and energy consumption in FL training are made.} 
    
\end{itemize}
This work is organized as follows. Section~\ref{related work} presents \ck{relevant} work on non-iid data distribution and client selection strategies \ck{in FL. System model considerations are provided} in Section~\ref{system model}. Section~\ref{sec:clientselection}
introduces the statistical similarity metrics (\ref{metrics}), the client selection strategy (\ref{client selection}), and the energy consumption model (\ref{energy}) used in this work. 
\ck{Performance results pertaining to the evaluation
of similarity-based clustering against random selection are provided in Section~\ref{evaluation}. Section~\ref{conclusion} is reserved for conclusions and discussion of the path forward.}
%

%







\section{Background and Related Work}
\label{related work}

The disadvantages of having all clients \ck{participating} in the FL training have led several studies to investigate the development and improvement of techniques for selecting a portion of clients to join in the FL training. However, \ck{the design of an optimal selection strategy is not a  straightforward task, mainly due to} the detrimental effects of non-iid data distribution on the behavior of FL training
~\cite{sattler2019robust,li2022federated,ma2022state,Luo2022}. For example, an experimental study is presented in~\cite{li2022federated} that investigates six non-iid partitioning strategies for different FL algorithms. 
Non-iid distributions are shown to have a significant impact on the accuracy and stability of FL algorithms, and the authors propose the implementation of client subgroups in each FL round to
deal with the instability caused by random sample selection. In~\cite{Luo2022}, on the other hand, the authors use the distributions of the labels to deal with heterogeneity in a non-iid medical scenario and focus on computing the marginal distribution for each client as an optimization strategy. 

In an effort to 
optimize the communication between the clients and the server, 
different client selection strategies have been proposed~\cite{Wang2021,balakrishnan2022diverse,lu2023federated,song2023v2x,jiang2022adaptive, Fraboni2021,briggs2020federated}. In~\cite{Fraboni2021}, client selection relies on the sample size and various similarity methods. 
As such, clients with more samples are weighted the most, which, in turn, does not prevent the transmission of redundant information, rendering this approach inefficient. It is also mentioned that measuring client similarity enables better clustering and leads to better performance. The considered similarity refers to the representative gradient, i.e., the difference between the updated model of a client and the global model, and is not extended to cluster similar clients. 
In~\cite{briggs2020federated}, hierarchical clustering is implemented 
to group FL clients according to the similarity of their model weight updates compared to the global joint model. The authors apply only distance metrics to compute the similarity between clusters.

A contextual client selection framework is proposed  in~\cite{song2023v2x} for a vehicle-to-everything scenario. The solution comprises information sharing, traffic topology prediction, and client clustering at both data and network levels. The clients are grouped at the data level 
using their gradient as a similarity criterion.  In~\cite{lu2023federated}, the authors used a probabilistic deep learning model to create a personalized feature extractor integrated into each client. To estimate client similarity, the server sends the model of all other clients to each client to measure the discrepancy between the two distributions. Based on the resulting discrepancy, the server then
clusters the clients.
The gradient 
is also used in~\cite{Wang2021}
as an insight to estimate the skewness of the client, 
using Hoeffding's inequality. A dueling bandit is then used to select the clients with the lower skewness based on this estimate. 
In~\cite{balakrishnan2022diverse}, the authors select a subset of active clients in each FL round using the stochastic greedy algorithm, which requires knowing the gradients from these clients. For this reason, a round of local model updates is performed before sharing the updates with the server. 

Our study differs substantially from previous approaches. First, rather than focusing on developing a strategy for client selection that needs to be integrated into the training phase, we propose a solution focusing on the client side that can be more easily integrated into a federated system. This is because our solution is centered on providing a client selection strategy before the FL 
\paolo{procedure starts.}
Second, although some works have used similarity metrics to improve FL client selection, \ck{they do not provide a rigorous comparative assessment of the metrics themselves.
As such, this work aims to compare a broad set of similarity metrics to gain in-depth insights into the FL performance under different degrees of skewness in label distributions.} 
Lastly, 
it is essential to obtain more efficient and sustainable \paolo{FL} systems (i.e., without requesting more computational, storage, or transmit power) and, at the same time, maintain or reduce energy consumption. Thus, it may not be enough to reduce the number of FL rounds in training, especially when we do not quantify the impact on energy consumption. This is a fundamental aspect that none of the previous studies have addressed. To fill these gaps, in this work we investigate the use of similarity metrics in the optimization of FL training, using the clients' data distribution to find a trade-off between the number of rounds and the energy consumption during this process. Relevant aspects \ck{which}, to the best of our knowledge, have not been investigated in a \ck{similar} study before.





\section{System Model}
\label{system model}
\begin{figure}[htbp]
    \centering
    \includegraphics[width=\columnwidth]{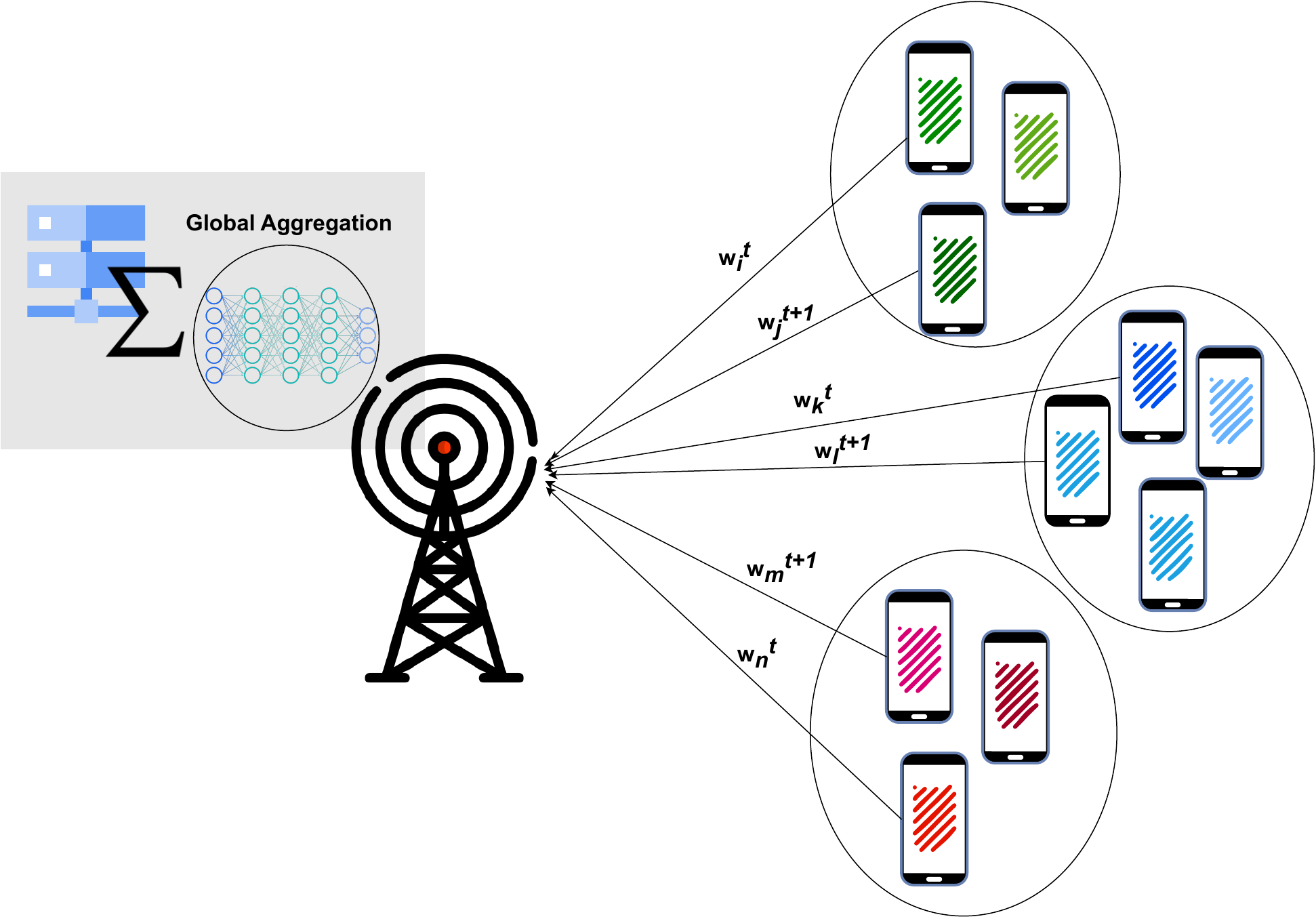}
    \caption{Overview of \ck{the considered problem setup in FL. Cluster formulation takes place with the aid of similarity metrics. At each communication round, client selection from clusters aims to harness the underlying heterogeneity of local training datasets.}}
    \label{fig:proposal}
    \vspace*{-0.5cm}
\end{figure}

In this work, we have adopted the FedAvg algorithm \ck{for distributed training \ff{\cite{mcmahan2017communication}}}, where \ck{a central} server aggregates a weighted average of the participating clients' model parameters. The weights are determined by the size of \ck{each} 
client's training dataset. We denote the number of clients as $N$ and the local training dataset \ck{of client $i$} as $D^i$ = $\{(\bm{x}_i,\bm{y}_i)\}$. \ck{Focusing on multi-label} classification \ck{as the learning task}, $\bm{x}_i$ stands for the data \ck{instances available at client $i$} and $\bm{y}_i$ represents the \ck{associated} labels. The local model of client $i$ in \ck{communication} round $t$ is \ck{expressed as} $w^t_i$, and the global model as $w_t$. Aggregation is performed after each communication round, 
as shown in Fig.~\ref{fig:proposal}.  Consequently, the participating clients optimize their local model before transmitting their model parameters to the server afterwards~\cite{li2022federated}.

Since the data collected by an underlying system \ck{are} usually non-iid \ff{\cite{ma2022state}}, we \ck{henceforth} focus on the use of similarity metrics \ck{to capture complex relationships 
among local datasets.}
\ck{These} metrics are essential to measure the similarity or dissimilarity between data instances. 
Although their use is primarily \ck{attributed} to transfer learning \ck{tasks}, \ck{such information measures} could be applied \ck{to a wide range of 
problems, such as feature extraction, information reuse, and multi-task learning, to name a few.}
In our \ck{case}, these \ck{statistical measures} are \ck{used} to form clusters, \ck{where 
client selection in each round of the federated training 
is performed in a way that promotes dissimilarity of selected clients}, as illustrated in Fig.~\ref{fig:proposal}.
%
This approach 
\ck{aims at a more efficient version of FedAvg scheme by avoiding  
the transmission of semantically} redundant information to the server.
\ck{As a result, given a certain accuracy threshold,}
the number of required rounds and the energy consumption to train the global model \ck{are expected to be reduced.} Consequently, the pressure on the \ck{underlying 
connectivity} can be alleviated
by saving network resources such as bandwidth and transmission power. 
\paolo{Moreover, non-selected clients at every round $t$ may not perform local training, thus saving computing power.}

\ck{We assume skewed label distribution that results in heterogeneity of the local models. The label distribution at each client is considered to be known. 
For clustering, we calculate the number of samples per label for each client, $n_{i,k}$, and we divide by the total number of samples $n_{i}$, as}
\begin{equation}
\label{eq:distributionprobability}
p_{i,k}= \frac{n_{i,k}}{n_i},
\end{equation}
\noindent where $p_{i,k}$ represents the distribution of label $k$ at client $i$. 
We perform this calculation
for all clients and we obtain \newff{$\boldsymbol{P}^{N\times K}$}, as 
\begin{equation}
\newff{\boldsymbol{P}} = \begin{bmatrix} p_{1,0} & p_{1,1} & \dots & p_{1,K-1}  \\ p_{2,0} & p_{2,1} & \dots & p_{2,K-1} \\  \vdots & \vdots & \ddots & \vdots  \\ p_{N,0} & p_{N,1} & \dots & p_{N,K-1} \end{bmatrix},
\label{eq:simMatrix}
\end{equation}
\noindent \ck{where $K$ denotes the total number of labels. Each row vector \newff{$\boldsymbol{p}_i$}, $i \in [1,N]$, in \newff{$\boldsymbol{P}$} is equivalent to the probability mass function of the labels for client $i$.} 


\section{Client Selection Using Similarity Metrics}
\label{sec:clientselection}
\subsection{Statistical Similarity Metrics}
\label{metrics}
This section presents the \ck{similarity} metrics considered in our approach and \ck{relevant} studies that have employed them in the past \ck{for other purposes.}

\paragraph{Cosine function} It \ck{estimates} the similarity between two vectors by \ck{measuring} the angle between them. \ck{The “closer” the two vectors are, the smaller the angle between them.} This \ck{widely used metric} is defined as
\setlength{\abovedisplayskip}{7pt}
\setlength{\belowdisplayskip}{7pt}
\begin{equation}
    \newff{cos_{i,j}} =\frac{\left \langle  \newff{\boldsymbol{p}_i}, \newff{\boldsymbol{p}_j}\right \rangle}{\left \| \newff{\boldsymbol{p}_i} \right \|\left \| \newff{\boldsymbol{p}_j}\right \|}, \quad \text{for} \quad i \neq j,
    \label{eq:cosine}
\end{equation}
\noindent where $\boldsymbol{p}_i,\boldsymbol{p}_j$ represent two different rows in \newff{$\boldsymbol{P}$}, \newff{and $\| \cdot \|$ denotes the $\ell^{2}$-norm (i.e., magnitude) of the vector}.
In~\cite{chen2022performance},
a cosine similarity function is used for original data recovery in the context of text transmission.

\paragraph{Mean Squared Error \ck{(MSE)}} This metric measures the average squared difference \ck{between each particular element in two different rows in \newff{$\boldsymbol{P}$}}. It is expressed as
\begin{equation}
    \newff{MSE_{i,j}} = \frac{1}{K}\sum_{k=1}^{K}\left ( p_{i,k} - p_{j,k} \right )^2, \quad \text{for} \quad i \neq j,
\end{equation}
\noindent where \ck{values closer to zero represent higher similarity.}
In \cite{sara2019image}, \ck{the metric is used for image quality assessment purposes.}

\paragraph{Euclidean distance} It is a \ck{widely used} 
metric that represents the shortest distance between two \ck{elements. Considering two different rows in} \newff{$\boldsymbol{P}$}, it is expressed as 
\begin{equation}
     \newff{D_{E_{i,j}}} = \left (\sum_{k=1}^{K}\left ( p_{i,k} - p_{j,k} \right )^2\right )^\frac{1}{2},  \quad \text{for} \quad i \neq j. 
\end{equation}
This metric was adopted in~\cite{briggs2020federated} to compute the similarity between clusters in a hierarchical clustering algorithm.

\paragraph{Manhattan distance} This metric computes the absolute distance between two \ck{elements in} \newff{$\boldsymbol{P}$}, 
defined as  
\begin{equation}
    \newff{D_{M_{i,j}}}= \sum_{k=1}^{K} \left | p_{i,k}-p_{j,k} \right |,  \quad \text{for} \quad i \neq j. 
\end{equation}
The metric was adopted in~\cite{briggs2020federated} for the same purpose as the Euclidean distance.

\paragraph{Chebyshev distance} This metric calculates the maximum of the absolute distance between two elements in \newff{$\boldsymbol{P}$}. It is defined as
\begin{equation}
    \newff{{D_C}_{i,j}}= \sum_{k=1}^{K}  \max_{k}\left | p_{i,k}-p_{j,k}\right |,  \quad \text{for} \quad i \neq j. 
\end{equation}
In~\cite{ozturk2021comparison}, the authors adopted this metric to detect whether images are similar or dissimilar in the context of dimensionality reduction.
\paragraph{Maximum Mean Discrepancy (MMD)}  It quantifies the distribution difference by computing the distance between the mean values of the instances in a reproducing kernel Hilbert space (RKHS)~\cite{zhuang2020comprehensive}. 
The calculation of MMD implies finding the RKHS function that maximizes the difference in the expectations $E(.)$ between two rows in \newff{$\boldsymbol{P}$}. 
In our case, we consider a linear kernel (i.e., $k(x,y)=\left \langle x,y\right \rangle$) and MMD can be calculated as 
\begin{equation}
\begin{split}
    MMD^2(\boldsymbol{p}_i,\boldsymbol{p}_j)=
    E_{\boldsymbol{p}_i}[k(x,x)]-2E_{{\boldsymbol{p}_i},{\boldsymbol{p}_j}}[k(x,y)]\\ \hspace*{-3cm}+E_{\boldsymbol{p}_j}[k(y,y)],  \quad \text{for} \quad i \neq j,
\end{split}   
\end{equation}
\noindent where $x\sim \boldsymbol{p}_i$, $y\sim \boldsymbol{p}_j$.
In~\cite{ghifary2014domain}, MMD is incorporated in a neural network model as a regularization technique to reduce the distribution difference between source and target domains in transfer learning. 

\paragraph{Kullback-Leibler (KL) divergence} It measures the statistical distance to minimize the divergence between two probability distributions, as 
\begin{equation}
    \newff{D_{KL_{i,j}}} =\sum_{k=1}^{K}p_{i,k}\log\frac{p_{i,k}}{p_{j,k}},  \quad \text{for} \quad i \neq j.
\end{equation}
This metric was adopted in~\cite{zhuang2015supervised} to measure the dissimilarity between domains as a mechanism to deal with the domain adaptation problem in transfer learning. 

\paragraph{Jensen-Shannon divergence} This metric is a symmetric version of the KL divergence, and aims to minimize the difference between two distributions~\cite{chen2017activity}. For $\boldsymbol{p}_i$, $\boldsymbol{p}_j\in \boldsymbol{P}$ with $i \neq j$, it is defined as
\begin{equation} 
    D_{JSD} (\boldsymbol{p}_i\parallel \boldsymbol{p}_j) =\displaystyle\frac{1}{2}\left ( D_{KL}(\boldsymbol{p}_i\parallel \boldsymbol{q}) + D_{KL}(\boldsymbol{p}_j\parallel \boldsymbol{q})\right ),
\end{equation}
\ck{where \newff{$\boldsymbol{q}$} is a mixed distribution defined as} $\boldsymbol{q}=\frac{1}{2}(\boldsymbol{p}_i + \boldsymbol{p}_j)$. This metric was adopted in~\cite{chen2017activity} to measure the similarity of features in an activity recognition problem.

\paragraph{Wasserstein distance} 
This metric 
measures the minimal effort required to reconfigure \newff{$\boldsymbol{p}_i$} in order to recover a distribution \newff{$\boldsymbol{p}_j$}, with $i$$\neq$$j$. 
In particular, we consider the \ck{1-Wasserstein} distance, expressed as
\begin{equation}
    W_1\left ( \boldsymbol{p}_i ,\boldsymbol{p}_j  \right ) =  \underset{\gamma \in \Gamma(\boldsymbol{p}_i,\boldsymbol{p}_j)}{inf} \int_{\mathcal{R} \times \mathcal{R}} \left | x-y \right |d\gamma\left ( x,y \right ),
    \label{eq:wass}
\end{equation}
\ck{where $\Gamma(\boldsymbol{p}_i,\boldsymbol{p}_j)$ denotes the set of probability measures
$\gamma$ on $\mathcal{R}\times\mathcal{R}$.
Elements $\gamma \in \Gamma(\boldsymbol{p}_i,\boldsymbol{p}_j)$ are called couplings of \newff{$\boldsymbol{p}_i$} and \newff{$\boldsymbol{p}_j$}, i.e., joint distributions on $\mathcal{R}\times\mathcal{R}$ with prescribed marginals \newff{$\boldsymbol{p}_i$} and \newff{$\boldsymbol{p}_j$}. Intuitively, Eq.~\eqref{eq:wass} implies that given a $\gamma \in \Gamma(\boldsymbol{p}_i,\boldsymbol{p}_j)$ and a pair of samples $(x, y)$, the value of $\gamma(x, y)$ reveals the proportion of \newff{$\boldsymbol{p}_i$}’s mass at $x$ that has to be transferred to $y$, in order to reconfigure \newff{$\boldsymbol{p}_i$} into \newff{$\boldsymbol{p}_j$}.}
In~\cite{lee2019sliced}, the authors adopted the Wasserstein distance to align the feature distribution domains and capture the dissimilarity between the outputs of task-specific classifiers.

\subsection{Client selection}
\label{client selection}
In this section, we integrate the similarity metrics into our client selection strategy. As shown in Algorithm~\ref{alg:named}, based on the selected similarity metric, the pairwise calculations between rows in \newff{$\boldsymbol{P}$} are performed.
Next, we leverage
k-medoids\footnote{K-medoids is publicly available in the Python library~\emph{scikit-learn-extra}:~\url{https://scikit-learn-extra.readthedocs.io/en/stable/modules/cluster.html#k-medoids}} as the clustering scheme to group the $N$ clients into clusters. 
For all possible cluster numbers $c\in [2,N-1]$, we calculate the silhouette value~\cite{Rousseeuw_1987} for client $i$ in order to determine the input value for k-medoids. The silhouette value is computed using
\begin{equation}
    s_c(i)=\frac{d_{\text{inter}}(i)-d_{\text{intra}}(i)}{max\{d_{\text{intra}}(i),d_{\text{inter}}(i)\}},
    \label{eq:silh}
\end{equation}
\noindent where $d_{\text{intra}}(i)$ denotes the mean intra-cluster distance, i.e., between the client $i$ and all other
clients in the same cluster and $d_{\text{inter}}(i)$ denotes the smallest mean inter-cluster distance, i.e., between the client $i$ and all other clients in any other cluster. 
The higher the silhouette value, the higher the probability of client $i$ being clustered in the correct group. 

\begin{algorithm}[h]
\small 
\caption{Applying similarity metrics to cluster the clients in a FedAvg algorithm.}
\label{alg:named}
\begin{algorithmic}[1]
 \renewcommand{\algorithmicrequire}{\textbf{Input: }}
 \renewcommand{\algorithmicensure}{\textbf{Output:}}
  \Require FedAvg inputs, metric $m \in \mathcal{M} = \{$cos, mse, mmd, $D_{kl}$, $D_{jsd}$, $W_1$, $D_{E}$, $D_{M}$, $D_{C}$, random\}, fraction of clients $\epsilon$  
 \Ensure The final model $w_t$
 \State Compute  $p_{i,k}= \frac{n_{i,k}} {n_i}$,  $\forall i \in N$, $k \in K$ 
 \State Construct $\boldsymbol{P}$ in Eq. \eqref{eq:simMatrix}
 \If {$m\neq$ random}
 \State Calculate pairwise similarities of $\boldsymbol{P}$ rows using Eqs. \eqref{eq:cosine}-\eqref{eq:wass}
\For {cluster number c$=2, \dots N-1$}
\State Calculate $s_c$ in Eq. \eqref{eq:silh}
\EndFor
\State Select $c$ with $\max\limits_{c}s_c$
and apply k-medoids for $c$ clusters 
  \EndIf 
 
 \hspace*{-1.1cm}\textbf{Server executes}:
  \Indent
  \State initialize $w_0$
  \For {each round $t =1,..., T_{\text{train}}$}
  \If{$m \neq$ random}
    \State $S_t \leftarrow$ (random set of $n$ clients in $c$ clusters)
    \Else
    \State $n\leftarrow max(\epsilon\cdot N,1)$ clients
    \State $S_t \leftarrow$ (random set of $n$ clients)
  \EndIf
  \EndFor
\EndIndent
\end{algorithmic}
\end{algorithm}
\vspace*{-0.01cm}



Once clustering using k-medoids is completed, the mapping of clients per cluster is available. Based on this information, the number of clients $S_t$ that will participate in the training rounds is determined.
On the other hand, when random selection is adopted, a predetermined number of clients are selected to join in each round (lines 15-16). 
After the client selection process, the conventional FedAvg algorithm follows. 

 
\subsection{Energy consumption}
\label{energy}
\ck{Besides efficient clustering, we also aim to derive energy consumption insights related to the FL process. \paolo{We focus on the computational energy, being the highest component in consumption, as concluded in our previous work~\cite{elia-ojcoms}.}
To this end, we evaluate \paolo{the computational} energy by considering a predefined set of clients $S_t$. In particular,} \paolo{the}
energy consumption for client $i$ is defined as
 \begin{equation}
 \label{eq:energy}
    e_i = P_{\text{hw},i} \cdot T_{\text{train},i}, 
 \end{equation}
where $P_{\text{hw},i}$ denotes the sum of the hardware power consumption of the GPU, RAM, and CPU for client $i$ during training time \newff{$T_{\text{train},i}$}.


\begin{figure*}[h!]
	\centering
	\subfigure[]{\includegraphics[width=0.4\textwidth]{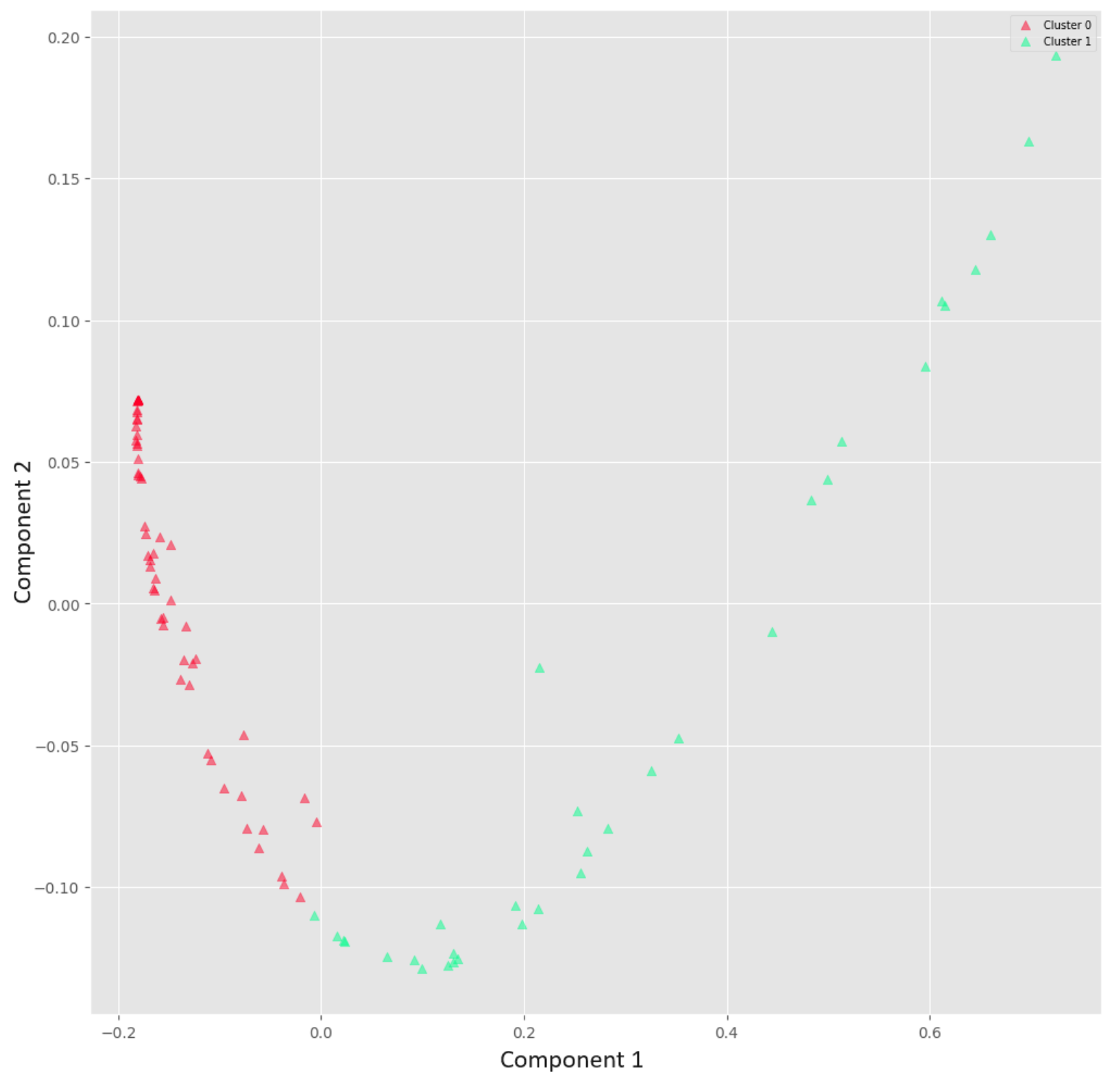}\label{fig:wasser}}
\hspace{0.5cm}
	\subfigure[]{\includegraphics[width=0.4\textwidth]{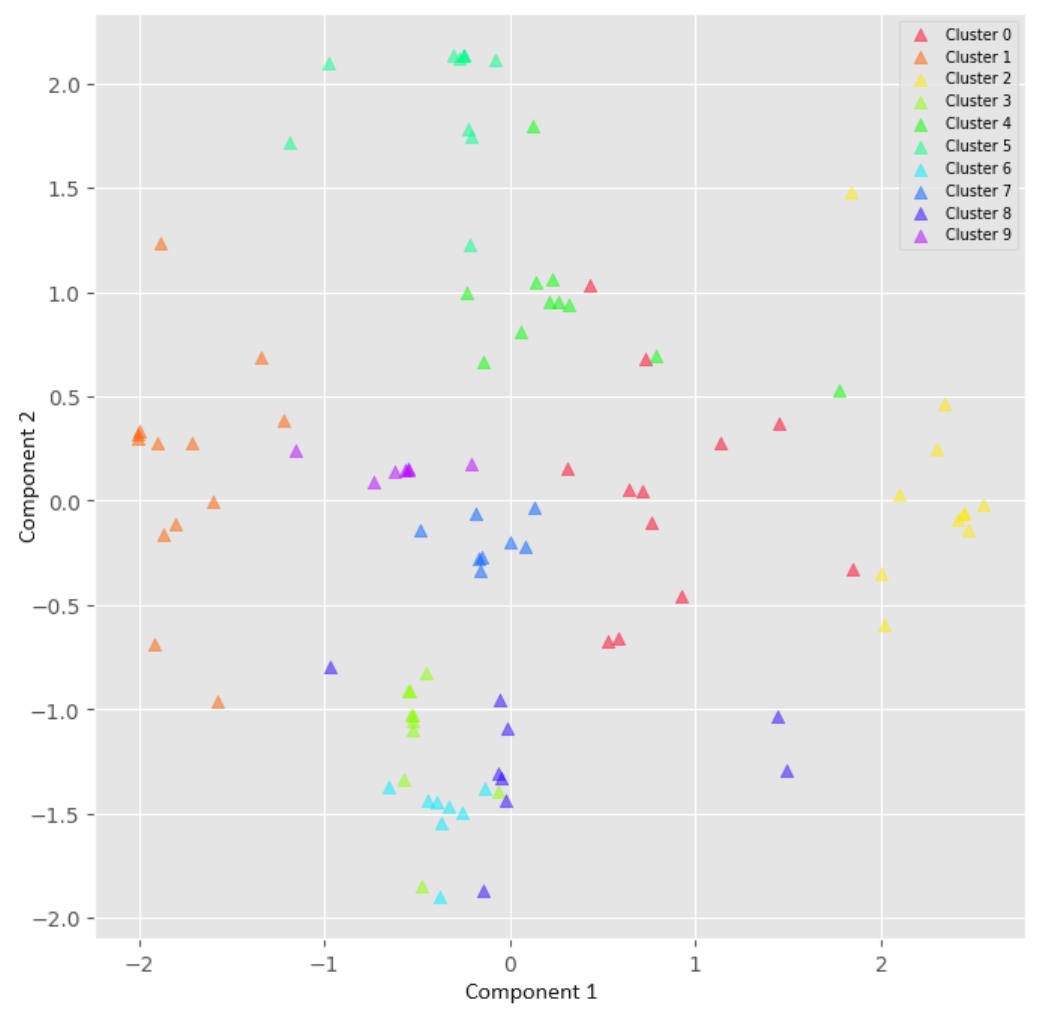}\label{fig:cheb}}
	\caption{Client clustering ($\beta=0.05$) with the aid of (a) 1-Wasserstein and (b) Chebyshev metrics. \newff{The principal component analysis (PCA) method was used as a dimensionality reduction technique to allow the mapping of features into two principal components.}}
	\label{fig:comp1}
 \vspace*{-0.4cm}
\end{figure*}

\section{Performance Evaluation}
\label{evaluation}

\subsection{Experimental setup} 

We implement and evaluate our \ck{approach} by \ck{performing} experiments on the well-known MNIST dataset~\cite{LeCun2010} of 28x28 images of handwritten digits, which comprises 60000 samples as a training set and 10000 samples as a test set. For all experiments, classification is performed with the aid of a convolutional neural network architecture comprising two 5x5 convolutional layers, followed by a 2x2 max pooling \ck{layer} and two fully connected layers with ReLU activation.
Experiments are performed on a server equipped with a 16-core Intel Xenon CPU and 2 NVIDIA GeForce RTX 3090 GPUs. We consider an FL system with \newff{$N$} = 100 clients. \ff{To generate a set of non-identical label distributions with different levels of skewness,} we apply
the Dirichlet distribution with a varying concentration parameter $\beta>0$~\cite{li2022federated}. 
In this way, we
can allocate a diverse proportion of \ck{samples}
of label $k$ to each client $i$.
The smaller the $\beta$, the more unbalanced the label distribution.


\subsection{Performance results}


For each similarity metric listed in
Section~\ref{metrics}, we evaluate the number of required rounds in the FL process and the standard deviation of the achieved accuracy for $3$ consecutive rounds, considering a predetermined accuracy threshold of 97\%. The computational energy during FL training is also assessed with the aid of CodeCarbon\footnote{CodeCarbon library is available at: \url{https://github.com/mlco2/codecarbon}} library~\cite{schmidt2021codecarbon} to estimate hardware power consumption according to Eq. \eqref{eq:energy}.
In addition, the baseline scheme of random selection with a varying number of participating clients $n \in [2, 5, 10, 15, 20, 25]$ is evaluated.
For all experiments, we consider 5 different seeds and calculate the average value of the obtained results.

\begin{table}[htpb]
\huge
\centering
\caption[Caption for LOF]{Similarity-based clustering vs random selection ($\beta=0.05$, accuracy=97\%) in FL training\footnotemark}
\resizebox{0.95\columnwidth}{!}{
\label{tab:beta005}
\begin{tabular}{|c|c|c|c|c|}
\hline
\multicolumn{1}{|c|}{\textbf{Metric}} & \multicolumn{1}{|C{3.1cm}|}{\textbf{Clients per round}} & \multicolumn{1}{C{3cm}|}{\textbf{Number of rounds}} & \multicolumn{1}{C{3.9cm}|}{\textbf{Energy consumption (Wh)}} & \multicolumn{1}{c|}{\textbf{Acc (std)}} \\ \hline
1-Wasserstein & \multicolumn{1}{|c|}{2}     & \multicolumn{1}{c|}{45.8}            & \multicolumn{1}{c|}{155.388} & \multicolumn{1}{c|}{0.003542}        \\ \hline
JS-divergence & \multicolumn{1}{|c|}{8.667} & \multicolumn{1}{c|}{39.333} & \multicolumn{1}{c|}{469.727}          & \multicolumn{1}{c|}{0.004051}      \\ \hline
KL-divergence & \multicolumn{1}{|c|}{8.667} & \multicolumn{1}{c|}{39.333} & \multicolumn{1}{c|}{471.221}          & \multicolumn{1}{c|}{0.004051}      \\ \hline
Euclidean &\multicolumn{1}{|c|}{10}    & \multicolumn{1}{c|}{45.8}            & \multicolumn{1}{c|}{329.746} & \multicolumn{1}{c|}{0.002496}          \\ \hline
Chebyshev &\multicolumn{1}{|c|}{10.2}  & \multicolumn{1}{c|}{45.8}            & \multicolumn{1}{c|}{328.94}  & \multicolumn{1}{c|}{0.001439}          \\ \hline
Manhattan &\multicolumn{1}{|c|}{10.4}  & \multicolumn{1}{c|}{45.8}            & \multicolumn{1}{c|}{400.298}          & \multicolumn{1}{c|}{0.003918}          \\ \hline
MSE &\multicolumn{1}{|c|}{10.4}  & \multicolumn{1}{c|}{45.8}            & \multicolumn{1}{c|}{427.103}          & \multicolumn{1}{c|}{0.005676}              \\ \hline
MMD &\multicolumn{1}{|c|}{10.4}  & \multicolumn{1}{c|}{45.8}            & \multicolumn{1}{c|}{428.517}          & \multicolumn{1}{c|}{0.005676}                \\ \hline
Cosine &\multicolumn{1}{|c|}{11}    & \multicolumn{1}{c|}{45.8}            & \multicolumn{1}{c|}{424.304}          & \multicolumn{1}{c|}{0.003945}             \\ \hline \hline
\multirow{6}{*}{
    \shortstack[l]{\textbf{Random} \\ \textbf{Selection}}}
 &\multicolumn{1}{|c|}{2}     & \multicolumn{1}{c|}{215.2}           & \multicolumn{1}{c|}{204.448}          & \multicolumn{1}{c|}{0.003261} \\ \cline{2-5}
&\multicolumn{1}{|c|}{5}     & \multicolumn{1}{c|}{113.40}          & \multicolumn{1}{c|}{377.386}          & \multicolumn{1}{c|}{0.003738}                \\ \cline{2-5}
&\multicolumn{1}{|c|}{10}   & \multicolumn{1}{c|}{87.8}          & \multicolumn{1}{c|}{563.378}          & \multicolumn{1}{c|}{0.006298}      \\ \cline{2-5}
&\multicolumn{1}{|c|}{15}    & \multicolumn{1}{c|}{63.80}           & \multicolumn{1}{c|}{621.135}          & \multicolumn{1}{c|}{0.001717}                    \\ \cline{2-5}
&\multicolumn{1}{|c|}{20}    & \multicolumn{1}{c|}{57.80}           & \multicolumn{1}{c|}{840.196}          & \multicolumn{1}{c|}{0.003119}                    \\ \cline{2-5}
&\multicolumn{1}{|c|}{25}    & \multicolumn{1}{c|}{49.4}            & \multicolumn{1}{c|}{726.045}          & \multicolumn{1}{c|}{0.002804}                    \\ \hline
\end{tabular}
}
\vspace*{-0.5cm}
\end{table}
\footnotetext{We ran the algorithm five times by changing the random seed and reporting the averaged classification accuracy.}

We first assess the FL performance
with similarity-based clustering and with random client selection in a highly heterogeneous scenario with $\beta\,$=$\,0.05$. 
The experimental results in Table~\ref{tab:beta005} reveal that the number of rounds required to reach convergence 
in the case of similarity-based clustering is lower compared to the random selection for all metrics.
It is worth noting that
the number of clients per round is not \textit{a priori} defined, but rather it is determined for each metric using Algorithm~\ref{alg:named} (line 8). A careful inspection of Table~\ref{tab:beta005} reveals that 
energy consumption can be remarkably reduced from $23.93\%$ to $41.61\%$ compared to the baseline random selection, for those similarity metrics achieving equivalent number of clients per round ($n=10$).

\begin{figure*}[ht!]
	\centering
	\subfigure[]{\includegraphics[width=0.4\textwidth]{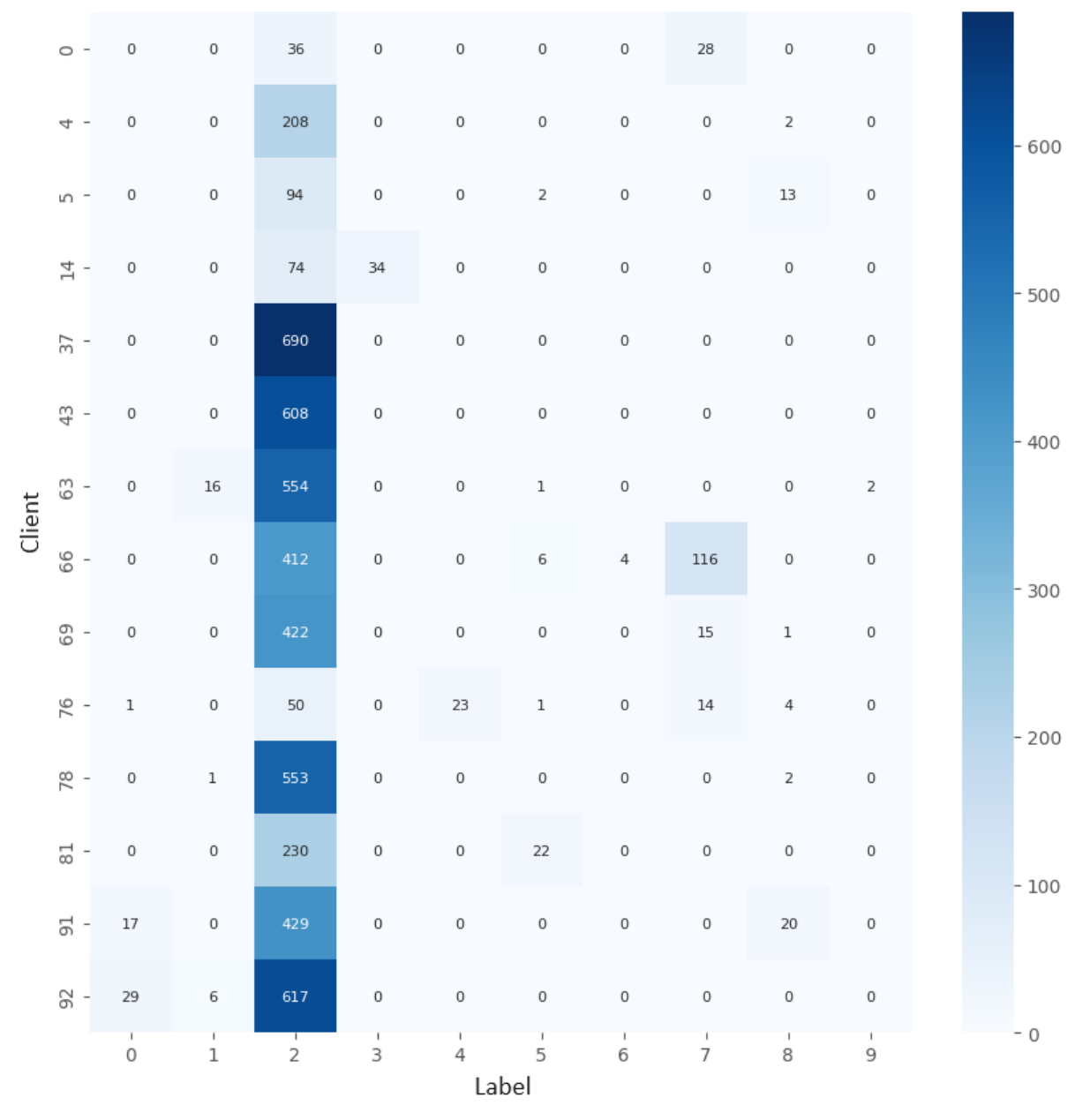}\label{fig:euclid}}
 \hspace{0.8cm}
	\subfigure[]{\includegraphics[width=0.4\textwidth]{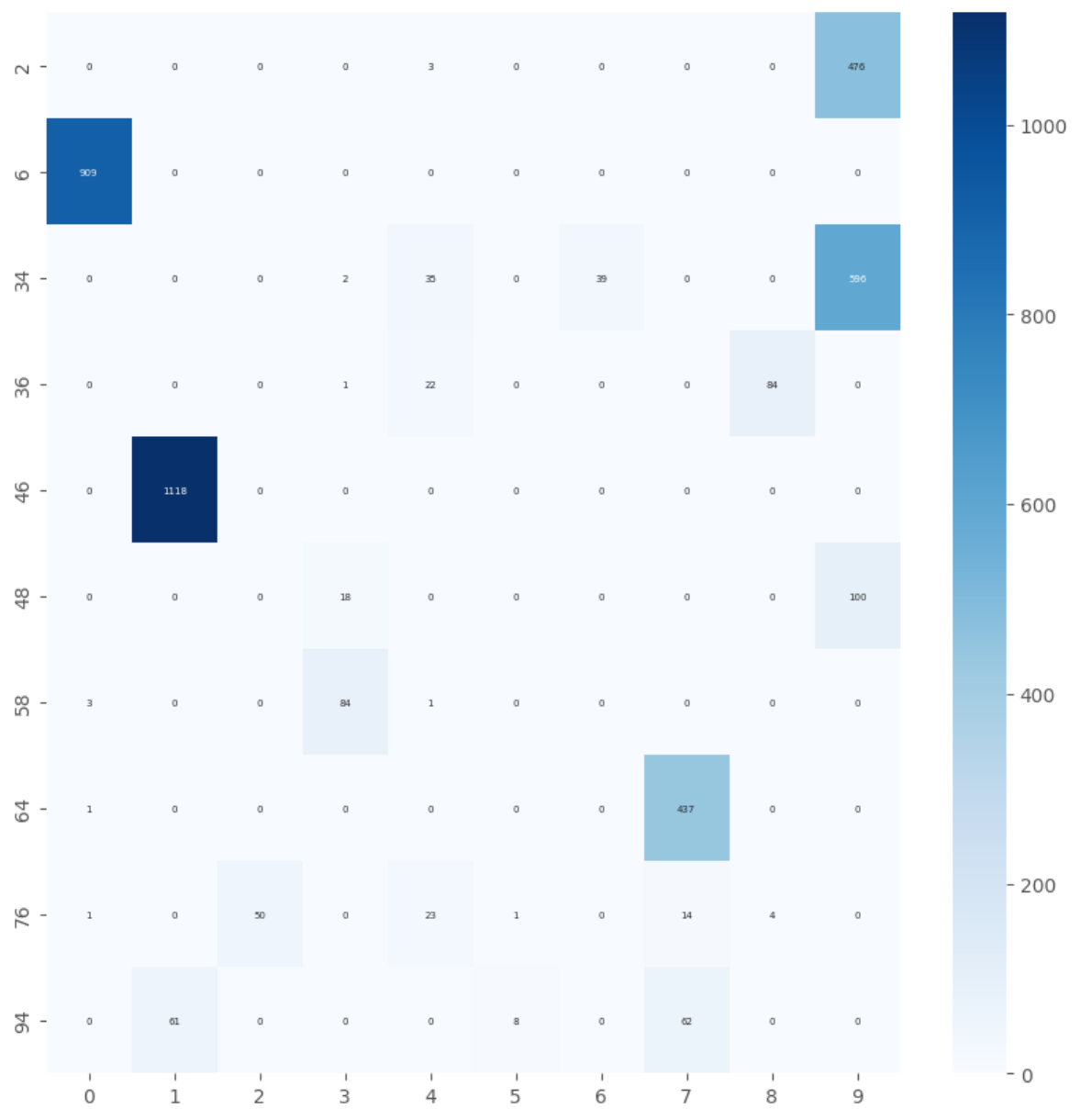}\label{fig:random-sec}}
	\caption{Clients of a single cluster ($\beta=0.05$, $n=10$) formed with (a) Euclidean distance metric (b) random selection. For random selection, we select the first out of $10$ clients in each FL training round.}
	\label{fig:comp2}
\end{figure*}

Notably, for the 1-Wasserstein distance, the number of required rounds for convergence is $\geq\,$$4.5$ times lower than in random
selection for the same number of participating clients ($n\,$=$\,2$). As illustrated in Fig.~\ref{fig:wasser}, a clear distinction between the two Wasserstein-based clusters can be achieved. 
In addition, the associated energy consumption in the training process remains at significantly lower levels compared to the random selection. 
On the other hand, clustering based on the Chebyshev similarity metric results in rather overlapping clusters\footnote{Note that similar behavior is observable also when using the other similarity metrics under study.}, as depicted in Fig.~\ref{fig:cheb}. \ck{The well-separated clusters under the 1-Wasserstein metric can be attributed to the fact that the Wasserstein space is able to subtly capture the geometry 
of the domain of the distributions in \newff{$\boldsymbol{P}$}~\cite{Panaretos_2019}.}





To demonstrate the
efficacy of similarity-based clustering, 
Fig.~\ref{fig:euclid}
depicts the resulting client agglomeration within a single cluster when Euclidean distance is applied for $n\,$=$\,10$. It can be observed
that similarity-based clustering allows grouping together those clients having samples predominantly of label $2$, including the client having only $36$ samples (client \#0). 
On the contrary, for random client selection, no relevant criterion for the label distribution is applied, resulting in an unstructured association of clients per cluster, as shown in Fig.~\ref{fig:random-sec} for $n\,$=$\,10$.

While the performance gains are noticeable for highly heterogeneous scenarios (i.e., low $\beta$), we henceforth explore the feasibility of similarity-based clustering in more homogeneous setups  with higher $\beta$. Table~\ref{tab:beta01} depicts the comparative performance outcomes for similarity-based clustering and random selection for $\beta\,$=$\,0.1$. In this case, it is noted that only Chebyshev and Manhattan metrics achieve superior performance in terms of both number of rounds and energy consumption compared to random selection for $n\,$=$\,10$. 
However, we remark here that even though the similarity-based clustering approach does not present such high-performance gains in more homogeneous data distributions, it does not require \textit{a priori} information on the number of clients to be selected at every round, as in the random approach. This feature represents an added value 
\ck{and poses elevated merit for fast FL deployments in realistic applications.} 
Finally, the performance gains of similarity-based clustering compared to random selection further vanish when $\beta\,$=$\,2$, as shown in
Table~\ref{tab:beta2}. It is noted, though, that such homogeneous settings are not common in real-world FL scenarios.

\begin{table}[t!]
\huge
\centering
\caption{Similarity-based clustering vs random selection ($\beta=0.1$, accuracy=97\%) in FL training.}
\label{tab:beta01}
\resizebox{0.95\columnwidth}{!}{
\begin{tabular}{|c|c|c|c|c|}
\hline
\multicolumn{1}{|c|}{\textbf{Metric}} &
\multicolumn{1}{|C{3.1cm}|}{\textbf{Clients per round}} & \multicolumn{1}{C{3cm}|}{\textbf{Number of rounds}} & \multicolumn{1}{C{4.1cm}|}{\textbf{Energy Consumption (Wh)}} & \multicolumn{1}{c|}{\textbf{Acc (std)}}  \\ \hline
 1-Wasserstein &
\multicolumn{1}{|c|}{2}    & \multicolumn{1}{c|}{127.2} & \multicolumn{1}{c|}{97.725}  & \multicolumn{1}{c|}{0.004085}        \\ \hline
Euclidean &
\multicolumn{1}{|c|}{9.8}  & \multicolumn{1}{c|}{34.8}  & \multicolumn{1}{c|}{139.428} & \multicolumn{1}{c|}{0.002823}           \\ \hline
Chebyshev &
\multicolumn{1}{|c|}{10}   & \multicolumn{1}{c|}{33.6}  & \multicolumn{1}{c|}{117.513} & \multicolumn{1}{c|}{0.003242}          \\ \hline
MSE &
\multicolumn{1}{|c|}{10}   & \multicolumn{1}{c|}{36.6}  & \multicolumn{1}{c|}{157.028} & \multicolumn{1}{c|}{0.004805}                \\ \hline
MMD &
\multicolumn{1}{|c|}{10}   & \multicolumn{1}{c|}{36.6}  & \multicolumn{1}{c|}{157.606} & \multicolumn{1}{c|}{0.004805}                \\ \hline
Manhattan &
\multicolumn{1}{|c|}{10.2} & \multicolumn{1}{c|}{31.4}  & \multicolumn{1}{c|}{122.535} & \multicolumn{1}{c|}{0.002134}          \\ \hline
Cosine &
\multicolumn{1}{|c|}{10.2} & \multicolumn{1}{c|}{31.4}  & \multicolumn{1}{c|}{140.241} & \multicolumn{1}{c|}{0.001658}        \\ \hline
JS-divergence &
\multicolumn{1}{|c|}{12.2} & \multicolumn{1}{c|}{60.8}    & \multicolumn{1}{c|}{149.1338} & \multicolumn{1}{c|}{0.003624}      \\ \hline
KL-divergence & \multicolumn{1}{|c|}{12.2} & \multicolumn{1}{c|}{60.8}    & \multicolumn{1}{c|}{164.8476} & \multicolumn{1}{c|}{0.003624}     \\ \hline \hline
\multirow{4}{*}{
   \shortstack[l]{\textbf{Random} \\ \textbf{Selection}}} &
\multicolumn{1}{|c|}{2}    & \multicolumn{1}{c|}{106}   & \multicolumn{1}{c|}{83.027}  & \multicolumn{1}{c|}{0.002556}  \\ \cline{2-5}
&\multicolumn{1}{|c|}{5}    & \multicolumn{1}{c|}{58.2}  & \multicolumn{1}{c|}{116.507} & \multicolumn{1}{c|}{0.005039}                    \\ \cline{2-5}
&\multicolumn{1}{|c|}{10}   & \multicolumn{1}{c|}{34}    & \multicolumn{1}{c|}{133.574} & \multicolumn{1}{c|}{0.002512}                    \\ \cline{2-5}
&\multicolumn{1}{|c|}{15}   & \multicolumn{1}{c|}{31.8}  & \multicolumn{1}{c|}{184.572} & \multicolumn{1}{c|}{0.002197}                    \\ \hline
\end{tabular}
}
\vspace*{-0.3cm}
\end{table}

\begin{table}[h]
\huge
\centering
\caption{Similarity-based clustering vs random selection ($\beta=2$, accuracy=97\%) in FL training.}
\label{tab:beta2}
\resizebox{0.95\columnwidth}{!}{
\begin{tabular}{|c|c|c|c|c|}
\hline
\multicolumn{1}{|c|}{\textbf{Metric}} &
\multicolumn{1}{|C{3.1cm}|}{\textbf{Clients per round}} &
  \multicolumn{1}{C{3cm}|}{\textbf{Number of rounds}} &
  \multicolumn{1}{C{4.1cm}|}{\textbf{Energy Consumption (Wh)}} &
  \multicolumn{1}{c|}{\textbf{Acc (std)}} \\ \hline
1-Wasserstein & \multicolumn{1}{|c|}{2}      & \multicolumn{1}{c|}{14}     & \multicolumn{1}{c|}{11.712} & \multicolumn{1}{c|}{0.003469}    \\ \hline
JS-divergence & \multicolumn{1}{|c|}{4}  & \multicolumn{1}{c|}{15.8}     & \multicolumn{1}{c|}{28.468} & \multicolumn{1}{c|}{0.002598}  \\ \hline
KL-divergence &\multicolumn{1}{|c|}{4}  & \multicolumn{1}{c|}{15.8}     & \multicolumn{1}{c|}{28.707} & \multicolumn{1}{c|}{0.002598}   \\ \hline
MSE&\multicolumn{1}{|c|}{20}     & \multicolumn{1}{c|}{8.6}  & \multicolumn{1}{c|}{76.105} & \multicolumn{1}{c|}{0.001238}             \\ \hline
MMD&\multicolumn{1}{|c|}{20}     & \multicolumn{1}{c|}{8.6}  & \multicolumn{1}{c|}{77.53} & \multicolumn{1}{c|}{0.000872}            \\ \hline
Chebyshev & \multicolumn{1}{|c|}{22}     & \multicolumn{1}{c|}{8.6}  & \multicolumn{1}{c|}{88.689} & \multicolumn{1}{c|}{0.001262}       \\ \hline
Manhattan& \multicolumn{1}{|c|}{24}     & \multicolumn{1}{c|}{10.4} & \multicolumn{1}{c|}{101.065} & \multicolumn{1}{c|}{0.001792}       \\ \hline
Euclidean & \multicolumn{1}{|c|}{24.8} & \multicolumn{1}{c|}{8.6}  & \multicolumn{1}{c|}{97.8324} & \multicolumn{1}{c|}{0.001689}      \\ \hline
Cosine & \multicolumn{1}{|c|}{26.6}     & \multicolumn{1}{c|}{9}  & \multicolumn{1}{c|}{88.689} & \multicolumn{1}{c|}{0.001262}        \\ \hline \hline
\multirow{6}{*}{\shortstack[l]{\textbf{Random} \\ \textbf{Selection}}} & \multicolumn{1}{|c|}{2} &
  \multicolumn{1}{c|}{12.6} &
  \multicolumn{1}{c|}{11.591} &
  \multicolumn{1}{c|}{0.003823}  \\ \cline{2-5}
&\multicolumn{1}{|c|}{5}      & \multicolumn{1}{c|}{8.4}  & \multicolumn{1}{c|}{19.098} & \multicolumn{1}{c|}{0.003137}                \\ \cline{2-5}
&\multicolumn{1}{|c|}{10}     & \multicolumn{1}{c|}{8.4}  & \multicolumn{1}{c|}{38.108}  & \multicolumn{1}{c|}{0.001083}                \\ \cline{2-5}
&\multicolumn{1}{|c|}{15}     & \multicolumn{1}{c|}{7.8}  & \multicolumn{1}{c|}{53.406}  & \multicolumn{1}{c|}{0.002364}                \\ \cline{2-5}
&\multicolumn{1}{|c|}{20}     & \multicolumn{1}{c|}{8}      & \multicolumn{1}{c|}{72.672} & \multicolumn{1}{c|}{0.001717}                \\ \cline{2-5}
&\multicolumn{1}{|c|}{25}     & \multicolumn{1}{c|}{8}      & \multicolumn{1}{c|}{92.24}   & \multicolumn{1}{c|}{0.001712}                \\ \hline
\end{tabular}
}
\end{table}

\section{Conclusion and Future work}
\label{conclusion}
\ff{In this paper, we proposed a similarity-based client selection approach to reduce the transmission of semantically redundant data. To this end, we incorporated nine statistical similarity metrics in the FL client clustering process.} 
\ck{Experimental results reveal that the more heterogeneous the local client data are, the more effective our method is, capitalizing on the lower number of required rounds and reduced energy consumption in FL training compared to random selection.}
\ck{In the path forward, we will direct our efforts towards integrating into our approach the quality of communication links between the participating clients and the server.}

\bibliographystyle{IEEEtran}
{\footnotesize
\bibliography{bibliography}}

\end{document}